\documentclass[useAMS,usenatbib]{mn2e}
\usepackage{epsfig}

\def\lcdm{$\Lambda$CDM }

\def\lcdmDot{$\Lambda$CDM. }

\def\lcdmCom{$\Lambda$CDM, }

\def\lcdmKet{$\Lambda$CDM) }

\newcommand{\mJy}{\mbox{$\tx{mJy}$}}

\newcommand{\hMpc}{\mbox{$h^{-1}$ Mpc} }
\newcommand{\hMpcKet}{\mbox{$h^{-1}$ Mpc)} }

\newcommand{\hMpcInv}{\mbox{$h\: \tx{Mpc}^{-1}$} }
\newcommand{\hMpcInvDot}{\mbox{$h\: \tx{Mpc}^{-1}$.} }
\newcommand{\hMpcInvCom}{\mbox{$h\: \tx{Mpc}^{-1}$,} }
\newcommand{\hMpcInvThree}{\mbox{$h^{3}\: \tx{Mpc}^{-3}$} }

\newcommand{\hMpcCom}{\mbox{$h^{-1}$ Mpc,} }
\newcommand{\hMpcDot}{\mbox{$h^{-1}$ Mpc.} }
\newcommand{\hmsun}{\mbox{$h^{-1}$ $M_{\odot}$} }
\newcommand{\hmsunKet}{\mbox{$h^{-1}$ $M_{\odot}$)} }
\newcommand{\hmsunCom}{\mbox{$h^{-1}$ $M_{\odot}$,} }
\newcommand{\hmsunDot}{\mbox{$h^{-1}$ $M_{\odot}$.} }

\newcommand{\kmsTwoDot}{\mbox{km$^{2}$ s$^{-2}$.} }

\newcommand{\PSCZ} {\emph{PSCz}\ }
\newcommand{\UZC} {\emph{UZC}\ }

\def\la{\mathrel{\hbox{\rlap{\hbox{\lower4pt\hbox{$\sim$}}}\hbox{$<$}}}}
\def\ga{\mathrel{\hbox{\rlap{\hbox{\lower4pt\hbox{$\sim$}}}\hbox{$>$}}}}

\def\gsim{\ga}

\newcommand{\bc}{\begin{center}}
\newcommand{\ec}{\end{center}}

\newcommand{\be}{\begin{equation}}
\newcommand{\ee}{\end{equation}}

\newcommand{\vechm}[1]{ \overrightarrow{\bf #1} }

\newcommand{\tx}[1] {\rmn{#1}}

\title{Constraining CMB-consistent primordial voids with cluster evolution}
\author[Hugues Mathis, Joseph Silk, Louise M. Griffiths and Martin Kunz]
	{H.~Mathis$^1$\thanks{Email: hxm@astro.ox.ac.uk}, 
	J.~Silk$^1$, L.~M.~Griffiths$^2$ and M.~Kunz$^3$ 	
	\\	
        $^1$Astrophysics, University of Oxford, Denys Wilkinson Building, Keble Road, Oxford OX1 3RH, UK\\
        $^2$Astrophysics, UNSW, Sydney, NSW 2052, Australia \\	
        $^3$Astronomy Centre, University of Sussex, Brighton BN1 9QJ, UK \\	
	}
	
\pagerange{\pageref{firstpage}--\pageref{lastpage}}

\begin{document}

\maketitle

\label{firstpage} 


\begin{abstract}

Using cosmological simulations, we make predictions for the 
distribution of clusters in a plausible non-gaussian model 
where primordial voids nucleated during inflation act together with 
scale-invariant adiabatic gaussian fluctuations as seeds for the 
formation of large-scale structure. The parameters of the void network  
are constrained by the cosmic microwave background (CMB) fluctuations 
and by the abundance and size of the large empty regions seen in 
local galaxy redshift surveys. The model may account for the 
excess of  CMB temperature anisotropy power measured on cluster 
scales by the Cosmic Background Imager (CBI). We show that the 
 $z=0$ cluster mass function differs little from predictions for a 
standard $\Lambda$CDM cosmology with the same $\sigma_8$, 
but that the evolution of the mass function at $z\sim1$ is slower 
than in a gaussian model. Because massive clusters form much earlier 
in the "void" scenario, we show that future integrated number counts 
of SZ sources and simple statistics of strong lensing will provide additional 
constraints on this non-gaussian model. 
 
\end{abstract}




\section[]{Introduction}
\label{sec:Intro}
The emergence of the $\Lambda$CDM scenario as the standard model for 
both the evolution of the cosmological background and the development of large-scale structure 
has recently received dramatic confirmation from the \emph{WMAP} microwave background experiment \citep{Spe03} and
from the large-scale distribution of galaxies \citep{Peac01,Teg03a}. 
 
In this picture of minimal complexity, adiabatic gaussian fluctuations in the energy density  
with scale--invariant power law power spectrum generated in the early Universe 
are stretched during inflation to astrophysically relevant scales. Gravitational instability is 
then responsible for amplifying the resulting overdensities and for  the collapse 
of structure. In  hierarchical models like cold dark matter (CDM), small objects collapse first 
and the more massive clusters form relatively recently. Although 
successful in many aspects, the ability of the model to reproduce 
 the  mass function of satellites of the Milky-Way \citep{Tu02,Sto02}, the rotation curves 
of LSB galaxies \citep{McG03}, or the bulk properties of the stellar populations 
of massive ellipticals \citep{Pee01} is still uncertain, even if recent hints 
towards tilt and/or running of the primordial spectrum index \citep{Pei03}
might help resolve some of the discrepancy on small scales.

On larger scales, the voids seen in the nearby galaxy distribution \citep{ElAd00,Pee01,Hoy02a} or the still controversial 
large-scale features \citep{Broad90,Perc01,Fri03} justify the development of  more complex,  non-gaussian scenarios
where linear, two-point predictions are in agreement with observations of the CMB, with the hope that simulations  
of galaxy formation in different environments and of the Lyman-$\alpha$ forest  will  agree with observations   
at least at the level of $\Lambda$CDM (\citealt{Kau99,Spr00,Ma02,Cro02}, see also \citealt{Bosch03}).

We consider one such alternative to the concordance model, where primordial
bubbles of true vacuum that formed in a first-order phase transition during
inflation can survive to the present day and result in cosmological voids
\citep{La91,Lid91,Occ97,Occ94}.  This model is typically implemented in the
context of so-called 'extended inflation' \citep{La89}. With reasonable
values for the distribution of voids, \citet[hereafter G03]{Gri03} (see also
\citealt{Ba97,Ba00}) have shown that the angular power spectrum of CMB temperature
anisotropies in this primordial void model with $\Lambda$CDM-type
cosmological parameters fits the observations at $l\la800$.  The signature
of voids emerges at $l\ga 1000$ and, if $\sigma_{8}=0.9$, could account for the excess power 
seen by the Cosmic Background Imager (CBI) on cluster scales ($l\sim 2500$) 
if it is not due to an underestimate of the Sunyaev-Zel'dovich (SZ) (see fig. 3 of G03). 
While viable in terms of the temperature angular power spectrum, we note that
\citet{Cor01} derive constraints on the underdensity and volume fraction of
voids at recombination from the COBE-DMR three-point correlation function and
that \citet{Ba00} predict non-gaussian features in the CMB temperature
fluctuations that have yet to be tested against high-resolution maps. On galaxy scales, 
G03 note that their void model may naturally account for the large underdensities 
seen in the local galaxy distribution which may be difficult to explain 
within the gaussian \lcdm paradigm (\citealt{Pee01,Gott03}, however see \citealt{Ma02}), 
and use the observed typical local void radii and filling fractions 
to set two of the free parameters of their void model.

To assess this void model from another direction which could lead to simpler observational constraints, 
we simulate large-scale structure formation using collisionless dark matter simulations of a network 
of compensated voids embedded in a $\Lambda$CDM cosmology. Due to the strong non-gaussianity, 
using concordance values for $\Omega_{0}$,
$\sigma_{8}$ results in a present-day cluster mass function (MF) that departs from the gaussian case. 
Non-linear structures develop much earlier in the void model, 
the result both of gravitational instability in the 
compensating dark matter shells surrounding the voids and 
of the large-scale motions triggered by the nonlinear 
evolution of the voids + shells systems.  Integrating  
the cluster evolution up to $z\sim5$,  we show that simple number counts of SZ sources and optical 
depth to strong lensing are enhanced with respect to the \lcdm scenario and that such 
observations may easily rule out the void model.  Note that \citet{Am94} have compared the angular 
two-point correlation function and the scaling of higher-order moments 
of a series of models with primordial voids to available observations of galaxy clustering. 
Their approach was however semi-analytical and lacked dynamical evolution of the structures.  

This paper is organised as follows. In Section 2, we recall  
the parameters of the void distribution and the approximations we make, the same as G03. 
In Section 3, we discuss the set-up of the initial conditions, make simple checks and present results at $z=0$. We deal with the high-z  
cluster evolution in Section 4, and motivate our choice of observables. We conclude in Section 5.


\section[]{Phenomenological model}
\label{sec:Model}

\subsection[]{Parameters}
\label{sec:Model:Param}

We take the parameters of the fiducial void model of G03. The background is a flat, dark energy dominated cosmology with
h=0.7, $\Omega_0=0.3$, $\Lambda_0=0.7$ and we normalize the amplitude of the mass fluctuations so that $\sigma_{8}=0.9$. 
This set of parameters is close to the best-fit obtained from the combined \emph{WMAP} + SDSS data \citep{Teg03b}.  
We assume that fluctuations of the field driving inflation produce 
the usual gaussian adiabatic scale invariant perturbations, filtered as they reenter the horizon by a CDM 
transfer function.  Variants of extended inflation predict that the first bubbles to nucleate during the phase 
transition can reach cosmological sizes at recombination. The cumulative number density of the voids is taken to be :
\be
        N_{V}(>r)=A\;r^{-\alpha}
\ee
where $r$ is the physical void radius, $A$ a normalisation constant 
adjusted to match the present-day filling fraction of voids seen in 
galaxy redshift surveys, and the exponent $\alpha$ is related to 
the gravitational coupling $\omega$ of the inflaton if extended 
inflation can be described  with a Brans-Dicke formulation : 
\be
        \alpha=3+\frac{4}{\omega+1/2} 
\ee
Solar system experiments require $\omega > 3500$ \citep{Wi01} and 
we will take here  $\alpha=3$. The lower and upper cutoffs 
in the void radii $r_{\tx{min,}\;\tx{max}}$ are 
chosen to agree with redshift surveys. 
\citet{Pl01} and \citet{Hoy02a} measured the typical size of voids in the  
\PSCZ and \UZC surveys and found  that up to 
half of the volume of the universe is underdense  
with $\delta\rho_{\tx{gal}}/\rho_{\tx{gal}} \la -0.9$. 
The void radii they obtain  range from $10$ to $30\; \hMpc$ with an average of $\sim 15\; \hMpcDot$ 
Following G03 we assume $r_{\tx{min,}\;\tx{max}}=10$, 25  \hMpcDot  (The contribution 
to the CMB temperature anisotropy angular power spectrum on CBI 
scales is mostly due to the voids with the largest radii; an analysis constraining $r_{\tx{max}}$ 
 using the CMB data alone is in progress.)  Examples of physical motivations  
for the lower and upper cutoffs are that (1) 
on subhorizon scales, matter flows 
relativistically back into the voids after inflation 
during radiation domination, and
suppresses the growth of small voids and that (2) 
the tunnelling probability of inflationary bubbles is modulated 
through the coupling to another field, resulting in a maximum radius. 
(See \citealt{Occ97} for a model of extended inflation involving two scalar fields and 
which gives a lower cutoff.) Last, we take an observed $z=0$ void filling fraction of $f_{\tx{voids}}=40\%$.

\subsection[]{Approximations}
\label{sec:Model:Approx}

In this section, we briefly review the assumed profiles and 
spatial distribution of the voids. 

We neglect the contribution of baryons as our purpose is to obtain statistics for massive clusters, where the dynamics is  
controlled by the dark matter. As the CDM becomes non relativistic early in the expansion,  
it is expected to travel only minimally into the voids which reenter the horizon later and are of interest 
for structure formation. On the other hand, due to the tight coupling to photons, baryons will fill in voids  which are 
within the horizon before decoupling at the adiabatic sound speed \citep{Lid91}, 
before gravitation takes over from radiation pressure. If at some point the mass of baryons inside the voids reaches a substantial  fraction of the mass of the shell, 
the growth of the void radius will be slower than expected for a fully empty, compensated region. 
We simply mention that the  precise dynamics of void filling by the baryons is complex  and needs further study when  
comparing for instance the high-redshift distribution of voids with the clustering of the Lyman-$\alpha$ forest, and that 
the precise density profile of the voids depends on the physics of reheating and of the subsequent filling by baryons. 

Smooth void density profiles have been considered in analytical work: rounded step functions \citep{Hof83,Mar90}, 
exponentials \citep{Haus83} or periodic functions \citep{Bac98}. 
However, with the exception of \citet{Re91} who also use a smooth initial profile,  numerical 
simulations of void evolution using particles often approximate voids as  ``top-hat'' underdensities \citep[hereafter W90, R00]{Dub93,Whi90,RB00}.   
These authors do not consider any particular physical model, but rather use generic templates of single voids or void networks. 
Here, to follow G03 we suppose that the voids at decoupling are spherical ``top-hat'' underdensities with  
$\delta\rho_{\tx{DM}}/\rho_{\tx{DM}}= -1$ surrounded by a thin compensating shell of dark matter.  

In an EdS cosmology, the shell density profile can be exactly derived from 
the self-similar solution of the evolution of a spherical underdensity \citep[hereafter B85]{Bert85a}.   
We assume that all the matter swept up during the expansion of the void ends
up in the compensating shell around it. This behaviour results naturally from the expansion of an  underdense region in an EdS cosmology 
(where the thickness of the shell is very slowly growing, see B85), a very good approximation  to \lcdm when we start the simulations. 
Recall that G03 assume for simplicity an EdS cosmology to compute the void contribution to the angular power spectrum of the CMB anisotropies, 
and  add the result to the power spectrum of the concordance \lcdm model employed here. This makes sense, because (1) most of this contribution 
comes from  voids close to the last scattering surface (LSS) where \lcdm is similar to EdS and (2) the 
contribution to the power spectrum varies smoothly with $l$, reducing the small impact of the correction for angular diameter 
distance even further. 

As in G03 we ensure that the voids do not initially overlap, although there seems to be no physical motivation for such a restriction. 
 In practice, given the starting redshift we choose, our results depend very weakly on this hypothesis.  Finally, the positions 
of centers of the voids are initially uncorrelated. 



\section[]{Simulating the void network}
\label{sec:Simu}

\subsection[]{Initial conditions}
\label{sec:Simu:IC}

We focus on two collisionless simulations of side 200 $\hMpc$  using $128^3$ particles, 
carried out with the publicly available N-body tree-SPH code {\sc gadget} without hydrodynamics 
\footnote{\tt http://www.mpa-garching.mpg.de/$\sim$volker/gadget/index.html}.  The first simulation, called $G$, is 
 a gaussian $\Lambda$CDM model with the above parameters, 
the second, $V$, is the \lcdm + voids non-gaussian 
fiducial model of G03. Except for the total initial displacement field, 
all simulation parameters are similar in the two cases. 
The simulations employ a Plummer softening length $\epsilon=0.08$ \hMpc which 
was kept fixed throughout in comoving coordinates.

The size of  the box is a compromise between the necessity of having enough primordial voids covering the whole 
radius range at $z=0$, $[\,10 \; 25\,]\; \hMpc$, and the mass resolution. 
(The comoving radius range is $[\,3 \; 6.3\,]\;  \hMpc$ at the starting redshift $z_{\tx{init}}$.)  
Even independently of the halo resolving power, $N_{\tx{parts}}=128^3$ is a stringent lower 
limit to the number of particles as simulations need to (1) propagate  information about the smallest 
voids present at the starting redshift, (2) have sufficient initial power in the cosmological displacement field  with 
respect to the power due to shot noise as one approaches the Nyquist frequency of the particles 
not to alter significantly the formation of the smallest haloes one can resolve and (3) 
ensure that 2-body scattering effects are not important.
While condition (1) is easy to verify with our number of particles, (2) and (3) are a little more complex, 
as these issues can be amplified because we use non-gaussian initial conditions for $V$. 
In addition, because of the strong primordial non-gaussianity, the formation of structure may also depend on the type 
of the initial distribution of particles which is employed before applying the initial displacement 
field, \emph{i.e.} whether one starts from a grid, a random (Poissonian), 
or a glass \citep{Ba95,Whi96} distribution. In the case of a grid, 
shot noise is minimal but the mesh introduces a characteristic scale and three preferred directions. A fully 
random, Poissonian distribution has no preferred direction but shot noise is significant. 
As a compromise between the two, we have 
used a glass distribution, which does not have any preferred direction, 
but with intrinsic power spectrum rising as $P(k)\propto k^{4}$. 
We find haloes with a friends-of-friends algorithm \citep{Da85} with a $z=0$ linking length $b\,\overline{l}=0.164$ times  
the mean interparticle separation $\overline{l}$. At redshifts $z=1$, 2 and 3, we have used the usual EdS linking length 
parameter $\overline{b}=0.2$ since \lcdm behaves as an EdS cosmology at these epochs (this  
corresponds well to the $b\propto (\Delta_{\tx{c}}/\Omega)^{-1/3}$ scaling proposed by \citealt{Eke96}, see also \citealt{Je00}). 
We keep only the groups with more than 10 particles and 
the minimum total halo mass we can resolve is $M_{\tx{min}}=3.16\times10^{12} \hmsunDot$ 
To address the issue of the choice of the initial distribution, we have checked 
that using a grid rather than glass for the initial distribution of particles in $V$ 
does not change the $z=0$ MF of dark matter haloes. To deal with  
the impact of the level of shot noise compared to the amplitude of 
the initial cosmological perturbations, we 
verify below that the simulated $z=0$ MF of $G$ matches analytical results 
(this is sufficient as we will show that the initial power spectrum of $G$ is smaller than that of $V$ over all simulated scales). 
Finally, we have found the $z=0$ number of dark matter haloes of $V$ to depend on the starting redshift when 
using only $N_{\tx{parts}}=64^3$ particles, a signature of unphysical resolution effects, 
but to have converged with $N_{\tx{parts}}=128^3$ particles.

In  $G$, the initial displacement field $\vechm{d}$ is given by the usual Zel'dovich 
approximation applied to the CDM power spectrum, 
normalised to a present $\sigma_{8}=0.9$. We compute $\vechm{d}$ on a $128^3$ mesh. 
In $V$, it is the \emph{same} $\vechm{d}$ for particles outside any primordial void and the displacement predicted 
by the similarity solution of B85 taken in the EdS regime ($r \propto t^{4/5}$ where $r$ is the physical void radius and $t$ the time)
for all particles that fall within a primordial void. Given the assumptions of paragraph~\ref{sec:Model:Approx}, we put all particles falling in a void at its radius, 
 and assign them the radial  velocity of the expanding shell, following W90.  In that sense, our $V$ simulation would 
be close to that of a ``spontaneous creation'' of voids at $z_{\tx{init}}$.

To follow G03, we use the similarity solution of an EdS universe (in particular the EdS time elapsed from $z_{\tx{init}}$ to $z=0$) to 
compute the initial radii and shell velocities of the voids, although we run the simulations with a \lcdm background.  In doing so, 
we neglect differences expected with respect to EdS in the initial radii 
if voids were scaled back from $z=0$ using a theoretical  solution for a \lcdm cosmology. 
However, we verify below that this enables simulations to correctly reproduce 
 single $25$ \hMpc radius voids at $z=0$. Furthermore, 
the similarity scaling between void radius and shell velocity remains 
strictly valid at $z_{\tx{init}}$ if it is high enough as EdS is then a very good approximation to \lcdm (see B85). 

The starting redshift for simulations of cosmological models 
with gaussian initial conditions is usually set by requiring the maximum 
particle displacement to be  $\sim30$ percent of the mean interparticle distance. 
An upper limit on $z_{\tx{init}}$  results from condition (1) on $N_{\tx{parts}}$ and translates the lower limit on 
the amplitude of the power spectrum of the initial perturbations. To choose  $z_{\tx{init}}$ for their  
 simulations of a non-gaussian model with primordial voids, R00 ensure that 
$\Delta^{2}<0.15$ at the Nyquist frequency of the particles and that no shell 
crossing occurs when initially displacing the particles. This is feasible 
as they compute the displacement for all their particles only  from the Zel'dovich scheme 
applied to a linear density field. Their density field is the superposition of a gaussian field and a distribution of mildly underdense spherical 
regions with $\delta_{\tx{void}}\sim 0.1$. Because we start our simulations with compensated empty voids, 
we immediately probe the non-linear regime on void scales. As a consequence, the MF of haloes that we obtain at $z=0$ 
in the void model could depend on $z_{\tx{init}}$: it is possible that some dark matter haloes that our simulation can resolve 
 have formed before $z_{\tx{init}}$ by fragmentation of the void shells, and starting at $z_{\tx{init}}$ could ignore them. 
Therefore we conservatively start the simulations soon after decoupling with a high $z_{\tx{init}}=1000$, still satisfying condition (1). 
We have checked that the $z=0$ halo mass function in $V$ obtained from starting at $z_{\tx{init}}=1000$ is similar to that 
obtained from starting at $z_{\tx{init}}=3000$ (this is not true for a simulation of $V$ with only $N_{\tx{parts}}=64^3$ particles).  
Independently of the shape of the one-point PDF of the initial overdensity field, a worry from such a high starting redshift 
is that numerical integration of the equations of motion might suppress the growth of small-scale modes 
(see the introduction of \citealt{Sco98}).  Again, we use below the $z=0$ halo MF of $G$ to check that this does not affect our conclusions.

The initial gravitational potential $\Phi_{\tx{tot}}(\vechm{x})$ that we construct in the 
 simulations can be linked to CMB observations. On void scales in the simulation $V$, like the overdensity field, 
$\Phi_{\tx{tot}}$ will deviate from a realisation of a gaussian random field. 
As an example, \citet{Kom03} look for a signature of non-gaussianity in the \emph{WMAP} 
CMB temperature fluctuations maps  
using variants of the bispectrum and Minkowski functionals. They then constrain 
the relative contribution $f^2_{\tx{nl}}$ of second-order 
deviations from a gaussian field expected in single-field inflation models (e.g. \citealt{Gan94,Ve00}) 
to the standard deviation of the gravitational potential
 but on a scale larger than the typical size 
 the voids studied here would have on the LSS. Nevertheless, analysis of non-gaussianity 
on smaller scales will be possible soon with interferometric maps, and   
further constraints on the fiducial \lcdm + voids model will result 
from comparisons with the theoretical bispectrum obtained in simulated 
maps of CMB temperature fluctuations.

Here we adopt a simpler approach to the non-gaussian contribution 
to the gravitational potential and estimate the relative amplitude of the 
gravitational potential energy of the gaussian component with respect  to the total energy  
in the compensated voids.  All energies are computed in proper coordinates; 
note that the total (kinetic and gravitational) energy of a compensated void, 
including the shell, is always positive (see B85). 
Recall also that in an EdS universe $W$ is constant if the growth of the fluctuations is linear. We 
note $E_{\tx{v}}$ the total energy of the network of voids and shells in the $V$ simulation 
and $W_{\tx{g}}$ the total gravitational potential energy of the gaussian density fluctuations
as realised in the $G$ (or $V$) simulation. 
At $z_{\tx{init}}=1000$, we obtain analytically $E_{\tx{v}}=7.09\times 10^{22}$  
while $W_{\tx{g}}=-7.79\times 10^{22}$ \hmsun \kmsTwoDot In the remainder of this section, 
we will implicitly assume this unit when quoting energies.

Our analytical estimates show that at the starting redshift, the energy due to the voids is similar to 
that of gravitational potential of the gaussian density perturbations. In addition, the 
probability distribution function of the gravitational potential 
of a network of compensated voids over a homogeneous background 
is not gaussian (it is strongly positively skewed). As a result, we expect 
 $f^2_{\tx{nl}}$ in the expansion of \citet{Kom03} to reach values of order unity 
on the angular scale of the voids, but a precise calculation also separating kinetic 
and gravitational energy of the voids+shells network is left for future work. 

In practice, because of shot noise and of transients 
from the Zel'dovich approximation, it is not easy to obtain 
a precise value for the potential energy at high redshift directly 
from the particle distribution in the simulations (irrespective of a possible primordial non-gaussianity). 
We have checked that at $z\sim10$ the predicted and measured values of $W_{\tx{g}}$ 
agree within 20 percent.

At $z=0$, using the EdS similarity solution for the evolution of the network of voids and shells in a \lcdm background, 
we obtain $E_{\tx{v}}=4.45\times 10^{22}$, and \emph{assuming} the \lcdm 
linear growth of the gaussian density fluctuations to be valid up to the Nyquist 
frequency of the particles we get $W_{\tx{g}}=-6.07\times 10^{22}$. 
With this hypothesis of linearity in our whole box, 
the energy of the voids and shells drops at late times to 70 percent of the potential energy. 
In other words, from $z_{\tx{init}}$ to $z=0$ 
the analytical $E_{\tx{v}}$ decreases by more than the factor 
$D_{\tx{lin},\Lambda\tx{CDM}}/D_{\tx{lin,EdS}}(z_{\tx{init}} \rightarrow 0)< 1$ describing 
the evolution of $W_{\tx{g}}$.

Directly in the simulations, we measure the total gravitational potential 
 $W_{\tx{sim}}= -2.22\times 10^{23}$ in $V$ 
and $-1.85\times 10^{23}$ in $G$. The excess factor of 2-3 
between measurements and analytical predictions for $W_{\tx{sim}}$ in $G$ is due 
to the non-linear evolution of the density field. The lower 
$W_{\tx{sim}}$ measured in $V$ compared to $G$ can be a consequence 
of the slightly larger number of haloes that we find in $V$ at $z=0$. 
On the basis of the predicted $E_{\tx{v}}$ at $z=0$, 
one would expect $W_{\tx{sim}}$ to be higher in $V$ than in $G$. 
This is not measured, however, because as seen in $G$ 
non-linear effects determine the late-time value of $W_{\tx{sim}}$.

For completeness, the total kinetic energy measured in the simulations right after 
setting the initial conditions is $K_{\tx{tot}}=1.44\times10^{16}$ and $1.60\times10^{14}$ in the $V$ and $G$ simulations respectively 
($K_{\tx{tot}}=1.19\times10^{23}$ and $1.02\times10^{23}$ at z=0). The peculiar velocities assigned 
to dark matter particles constituting the void-compensating shells is responsible for the large excess of 
initial kinetic energy in $V$. At late times, this excess is washed out in the dominant contribution 
of the particles of virialized haloes.

\subsection[]{Consistency checks}
\label{sec:Simu:Tests}
  
To check the effect of periodic boundary conditions,  
we simulate with {\sc gadget}  the growth from $z_{\tx{init}}$ of a single void 
with final radius 25 \hMpc centered on a 100 \hMpc box without the 
gaussian part of the displacement field, in a EdS universe. We use the EdS similarity solution  
to scale back to $z_{\tx{init}}$.  At $z=0$, we find that the total mass enclosed inside 99 \% (98 \%) of the 
inner shell radius of $25$ \hMpc is 9 \%  (3\%) of that expected in a similar volume with mean density.  
The same test is repeated with the scaling method described above 
for simulations in a \lcdm background and we find a similar ``leaking'' mass fraction, see Fig.~\ref{fig:TestRad}.
 In both cases a large fraction this remaining mass is contributed by particles of clusters formed on the shell. 
Down to $z=0$, the particle distribution remains smooth beyond a $27$ \hMpc radius. 

\begin{figure}
\centering
\caption{Final particle distribution in a \lcdm 
background simulation (without the cosmological gaussian perturbations) of a 25 \hMpc - radius void centered 
on a 100 \hMpc box. The slice shown has thickness 
 10 \hMpcDot The radii of the inner and outer solid circles are respectively 
the comoving input at $z_{\tx{init}}$ (6.3 \hMpcKet and  25 \hMpcDot}
\label{fig:TestRad}
\end{figure}

We verified that the $z=0$ halo MF of the $G$ simulation gives results in agreement with the 
fitting formula of \citet[hereafter J01]{Je00} and that it is robust against changes in $z_{\tx{init}}$. 
(Using $N_{\tx{parts}}=32^3$ rather than $N_{\tx{parts}}=64^3$ or $128^3$ for the $G$ simulation results 
in an excess of low mass haloes if $z_{\tx{init}}=1000$, and using $N_{\tx{parts}}=64^3$ 
rather than $128^3$ for the $V$ simulation leads to a redshift dependence of the MF: 
a factor 3 more massive haloes than for $N_{\tx{parts}}=64^3$ and $z_{\tx{init}}=100$). 

The validity of using the Zel'dovich approximation when setting up the initial conditions
is checked on large scales by the growth of the largest modes of the simulation, in agreement 
with linear theory down to $z=0$. We have finally performed two other simulations 
of the void model, changing both the gaussian random perturbation field and the positions of the voids, 
and found results very similar to those for $V$.

\subsection[]{Results at $z=0$}
\label{sec:Simu:Resz0}

	Fig.~\ref{fig:Slice} shows the projected (2D) density of 
two slices of side 200 \hMpc and thickness 20 \hMpc cut at the same position 
through the $V$ and $G$ simulations (left and right panels respectively). 
Recall that the two simulations use the same initial gaussian displacement
field, which in the case of $V$ is combined with the displacement due to the primordial voids. 
The grey scale is the same in the two cases. Note the voids in the slice through $V$, 
apparent immediately below the centre of the picture and at the middle of the upper frame, 
together with a void network developing at the centre left.

\begin{figure*}
\begin{minipage}{160mm}
\centering
\caption{Left and right panels: projected 3D density at $z=0$ in slices cut at the same position 
through the  \lcdm + voids and gaussian (\lcdmKet  simulations respectively. 
The side and thickness are 200 and 20 \hMpcDot The \emph{gaussian} initial displacement  
field is the same in both simulations. The grey scale is the same for the two panels. 
Note on the left panel the voids right below the centre of the picture, at the middle of the upper frame, 
and a void network developing a ``honeycomb'' structure at the centre left.}
\label{fig:Slice}
\end{minipage}
\end{figure*}

\begin{figure*}
\begin{minipage}{140mm}
\centering
\epsfig{file=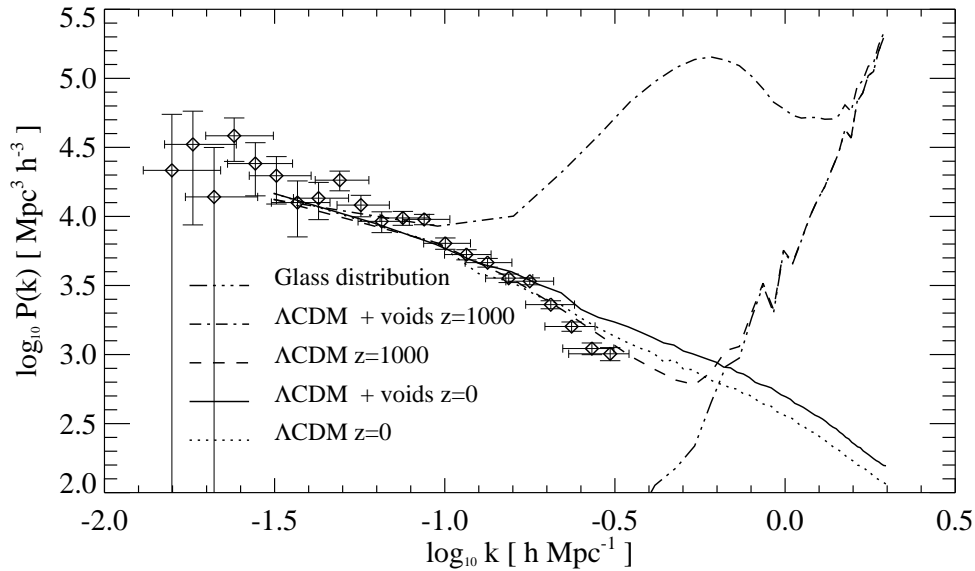,width=14cm}
\caption{Real-space matter overdensity power spectrum at $z=0$ of the gaussian 
(\lcdmKet and non-gaussian (\lcdm + voids) simulations :  dotted and
solid lines respectively. The dashed and 
dash-dotted lines show the same respective quantities at $z=1000$, 
after division by the linear growth factor for clarity. 
The dash-triple dotted line is the power spectrum 
of the raw glass file, for comparison. Note the signature of the voids 
in the initial conditions at $k\gsim0.1$ \hMpcInv and the 
stronger $z=0$ non-linear power in the void model. Diamonds with error bars show 
the real-space galaxy power spectrum from \citet{Teg03a}  (a more direct comparison between the \lcdm + voids simulation
 and the data would need a detailed model for galaxy bias).}
\label{fig:PS}
\end{minipage}
\end{figure*}

Fig.~\ref{fig:PS} gives the initial and $z=0$ real-space 
overdensity power spectra (dotted and solid lines for 
the $G$ and $V$ simulations respectively). 
Also shown are the power spectra of the unperturbed glass initial 
particle distribution (dash-triple dotted line), 
and of the initial conditions for the $G$ and $V$ simulations 
(dashed and dash-dotted lines). All spectra have been divided 
by the linear growth factor for clarity.  The diamonds with associated error bars show the real-space 
galaxy power spectrum as measured by \citet{Teg03a} from the DR1 of the SDSS survey.   

The imprint of the voids is significant at $z_{\tx{init}}$ at $k\gsim0.1 \hMpcInvCom$ 
as is clear when comparing the dash-dotted and dashed lines. 
At $z=0$, non-linear power is larger in the void model than in the \lcdm model, and departure 
from the evolved \lcdm power spectrum occurs as early as $k=0.1$ \hMpcInvDot Of course 
a precise comparison between the $V$ power spectrum and the data would need to correct for bias, 
a quantity that might show a different behaviour  in the \lcdm + voids model 
than in \lcdmDot If a precise assessment of galaxy bias falls beyond the scope of this work, 
we note that at $k=0.25$ \hMpcInvCom the dark matter bias 
of the \lcdm + voids model with respect to the \lcdm model 
is not more than 1.15. At the same scale, the bias of 
the evolved dark matter density field of the \lcdm + voids model with 
respect to the galaxy distribution reaches 1.3, a high but plausible value. 

\begin{figure*}
\begin{minipage}{140mm}
\centering
\epsfig{file=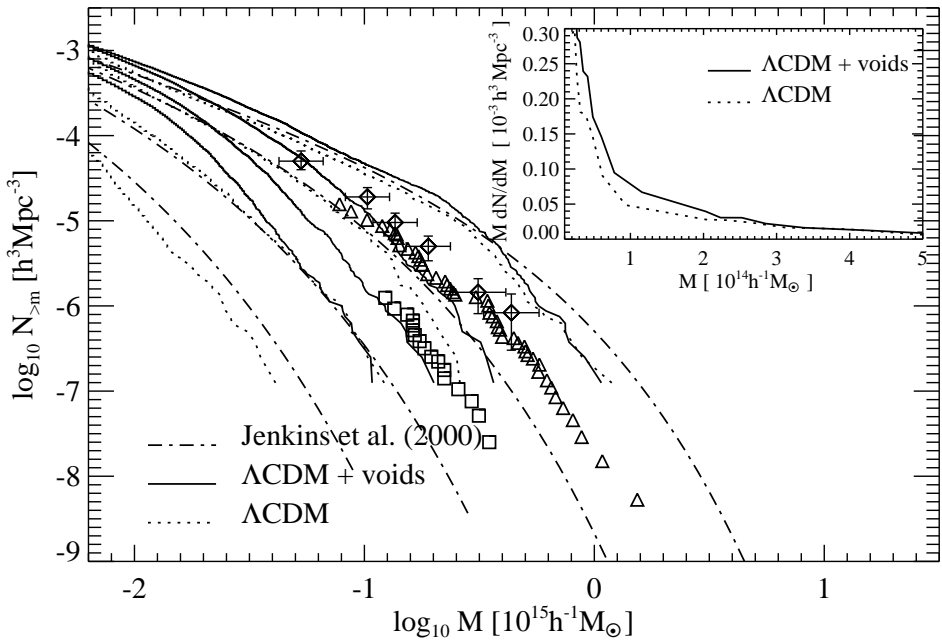,width=14cm}
\caption{Main panel: mass functions of the gaussian (\lcdmKet  and non-gaussian 
(\lcdm + voids) simulations: dashed and solid lines respectively, measured 
at $z=3$, 2, 1 and 0 from left to right. The dash-dotted line is the fitting formula 
of J01 to \lcdmDot The diamonds with error bars show the ``optical'' mass 
function that  \citet{Bah03a} obtain using the EDR of the SDSS. 
The triangles and squares give the cluster total mass functions at $z=0$ 
and $0.4<z<0.8$ respectively. They have been derived from the 
cluster baryon mass function constructed by \citet{Vik03,Voe03} 
using a combination of optical and X-ray data (their error bars are omitted for clarity). 
Insert: differential mass functions $\tx{d}\tx{N}/\tx{d}\ln{\tx{M}}$ of the gaussian (\lcdmKet  
and non-gaussian (\lcdm + voids) simulations: dashed and solid lines respectively, 
measured at $z=0$ up to $5\times10^{14}$\hmsunDot}
\label{fig:MF}
\end{minipage}
\end{figure*}

The main panel of Fig.~\ref{fig:MF} compares the halo MF 
measured at $z=3, 2, 1$ and $0$ from left to right in the $G$ and $V$ simulations 
(dotted and solid lines respectively) to the data and to the fitting formula of J01 
calculated for the parameters of the gaussian \lcdm model we have simulated   
(dash-dotted lines). While the $G$ simulation agrees well with J01,  
there are systematically more haloes in the $V$ simulation at $z=0$, 
for mass thresholds $M_{\tx{min}}\le5\times 10^{14} \hmsunDot$  
The excess is a factor 1.5 in the abundance. For higher mass thresholds, 
the abundance is similar in the two models. 
This is more clearly seen in the insert in Fig.~\ref{fig:MF}, which gives 
the differential halo MF measured around $2\times10^{14}$\hmsunDot
The excess of the non-gaussian model is apparent at all mass scales  
smaller than $3\times10^{14}$\hmsunDot For higher masses the two curves are identical. 
We find one halo with $M_{\tx{tot}}>10^{15} \hmsun$ 
in both the $G$ and $V$ simulations. (Recall that the gaussian part of the initial overdensity field is 
normalised to the same  present-day $\sigma_{8}=0.9$.)

\citet{Bah93} give $N_{>M}=2 \pm 1\,\times 10^{-6}$ \hMpcInvThree at $M=4\times 10^{14}$ \hmsunDot 
Even assuming their uncertainty factor of 1.3 in the mass of rich clusters, 
this normalisation falls significantly below both our $V$ and $G$ mass functions. 
 The present-day MF of the $V$ simulation is therefore only marginally 
consistent with this normalisation, but the level of disagreement is the same as 
that of a \lcdm model with concordance cosmological parameters and $\sigma_{8}=0.9$. 
Diamonds with error bars on Fig.~\ref{fig:MF} show the more recent data of \citet{Bah03a,Bah03b} 
who constrain the amplitude of mass fluctuations using the ``optical'' mass function 
of clusters selected from the SDSS EDR. They find a normalisation similar to that of 
\citet{Bah93}, with best-fit values $\sigma_{8}=0.9$ and $\Omega_{0}=0.19$. 
From a combination of X-ray and optical observations, \citet{Vik03,Voe03} 
derive constraints on the cosmological parameters $\Omega_{0}$, 
$\Lambda_{0}$ from the evolution of the cluster baryon mass 
fraction up to $z\sim0.5$, requiring $\sigma_{8}$ 
to be set by the observed amplitude of the $z=0$ baryon mass function. 
Their $z=0$ and $0.4< z < 0.8$ total mass functions for massive clusters 
are reported with triangles and squares respectively on Fig.~\ref{fig:MF}  
(we have omitted their error bars for clarity, and corrected 
for a weak mass dependence of the cluster baryon fraction, 
see \citealt{Vik03}). The $z=0$ mass function of \citet{Vik03} agrees with \citet{Bah03a}. 
At high redshift, the observed mass function has a lower normalisation 
than both the $G$ and $V$ $z=1$ mass functions, a repeat of the $z=0$ differences.

In fact, it is clear from Fig.~\ref{fig:MF} that the difference of the fiducial model of G03 with respect to \lcdm is 
much more significant in the intermediate and high-z mass functions. Unfortunately, measurements 
of the cluster mass function at high redshift $z\gsim1$ from X-ray observations are subject to uncertainties 
in, for instance, the redshift evolution of the L-T and M-T cluster scaling relations (see however \citealt{Voe03}).  
In the following section we will consequently use integrated SZ cluster counts and 
the abundance of giant arcs to probe the signature of the void model, rather than the cluster mass function.  

We conclude this section with two additional tests at $z=0$, first 
 changing the parameters of the void model and then 
comparing with two gaussian models.

We have simulated another primordial void model $V'$ of the same type as $V$ 
but with a much steeper spectrum for the distribution of initial void radii 
(using $R_{\tx{ min, max}}=10$, 40 \hMpcCom $\alpha=6$, $f_{\tx{voids}}=40\%$). 
G03 show that this model also agrees reasonably well  with high and low-$l$ measurements 
of the angular power spectrum of the CMB temperature fluctuations. The $V'$ $z=0$ cluster MF shows  
 a large excess compared to the $V$ MF (an order of magnitude more massive clusters and $\sigma_8\sim2$), so that this 
set of void parameters is directly ruled out. 

Second, it is necessary to check the impact of non-gaussianity by comparing to a model with 
 gaussian initial conditions but with an initial power spectrum similar to that measured in the void model 
right after setting the initial conditions. For this purpose, we have simulated two gaussian models $V_{1}$ and $V_{2}$ 
with the same cosmology and starting redshift as $V$ and $G$ but taking as 
initial power spectrum the dash-dotted line $P_{\tx{voids,init}}$ of Fig.~\ref{fig:PS}. 
$V_{1}$ and $V_{2}$ correspond to two different normalisations of the input power spectrum: for $V_{1}$ 
we have taken $P_{\tx{voids,init}}$ of Fig.~\ref{fig:PS} divided by the linear growth  
from $z=1000$ to $z=0$, while for $V_{2}$ we have normalised so that the simulation 
has $\sigma_{8}=0.9$ at $z=0$ (the adopted normalisation for $V_{1}$ 
results in $\sigma_{8}\sim4$ at $z=0$). We have found the $z=0$ cluster mass function 
of $V_{1}$ to largely exceed that of $V$ and $G$ (there is a factor of 5 
more haloes with masses $M_{\tx{tot}} > 4\times10^{14}\hmsunKet$ 
and that of $V_{2}$ to be abruptly cut at masses $M_{\tx{tot}} \sim10^{14}\hmsunDot$ 
In addition, the matter power spectra measured at $z=0$ in $V_{1}$ and $V_{2}$ 
retained the strong feature (``bump'')  seen in the initial conditions at $k_{\tx{void}}> 0.1 \hMpcInv$ 
as a step-like increase at $k_{\tx{void}}$ which was only little modified 
by the late-time non-linear evolution. This is clearly ruled out by the data. 
To summarise, the non-gaussianity of the primordial void model 
is necessary to approximately reproduce the $z=0$ observed 
cluster mass functions and mass power spectra: employing the initial 
power spectrum of the void model in a gaussian primordial 
density field results in a large mismatch to the data. 
The large peculiar velocities of the compensating shells surrounding 
the voids and associated to the scale $k_{\tx{void}}$ are not realised in a gaussian model with same initial power spectrum as $V$. 
As a result, the pattern at $k_{\tx{void}}$ in the power spectrum 
is more stable and persists longer in our $V_{1}$ and $V_{2}$ tests than in $V$, where large velocities 
may dilute/broaden the feature. The 
particular shape of the power spectrum resulting from the 
void network is not transposable to a gaussian initial probability function.


\section[]{Deriving observational constraints}
\label{sec:Obs}

In this Section, we make predictions for the thermal SZ effect and for simple statistics of strong lensing. We show 
that they differ substantially in the primordial void model from their values in a \lcdm cosmology. We note here that 
other observables like the cosmic shear or the clustering of the Lyman-$\alpha$ forest  could also bring 
out the presence of primordial voids, but they are more complex than the former and may be affected 
by biases due to the non-linear evolution of the power spectrum. 

\subsection[]{SZ source counts}
\label{sec:Obs:SZ}

\begin{figure}					
\centering
\epsfig{file=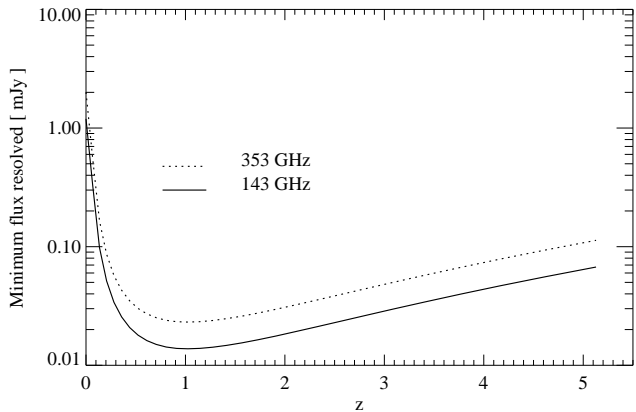,width=9cm}
\caption{143 and 353 GHz resolution limit for the cluster thermal SZ effect in the simulations as a function of redshift. 
The sensitivity of \emph{Planck} to the SZ effect is of order 30 mJy \citep{Bart01}, that of the 
planned Bonn-Berkeley \emph{APEX - SZ} survey will reach 3 mJy at 150 GHz.}
\label{fig:MinSZFlux}
\end{figure}

We first estimate the counts expected from the detection of the 
cluster thermal SZ effect, up to $z=5$.  \citet{Kay01} make detailed analytical predictions for the SZ number counts 
expected for the \emph{Planck} satellite using large simulations 
of cluster formation in gaussian CDM cosmologies. We follow the same approach but with simplifying assumptions.  
We  use  30 simulation outputs for $G$ and $V$ from $z=0$ to $z\sim5$.
Each dump has a comoving size 200 \hMpc and we find their DM halos 
above the minimum  mass threshold $M_{\tx{min}}$.  For simplicity we assume 
for each halo an isothermal profile for the gas (see, e.g., \citealt{Barb96}) 
with total mass $M_{\tx{gas}}=f_{\tx{b}}\times M_{\tx{tot}}$ where  $f_{\tx{b}}=0.13$ 
is the cosmic baryon fraction. Taking the clusters 
to be point sources, we compute the magnitude of the flux 
change observed against the CMB at  the \emph{Planck} satellite frequencies 
143 and 353 GHz, on both sides of the zero-point of the SZ thermal effect (217 GHz). 
Fig.~\ref{fig:MinSZFlux} shows the  minimum SZ flux 
that the simulations can resolve as a function of $z$. In the following, we show results for fluxes 
 $S_{\nu}> 10\; \mJy$. 


\begin{figure}
\centering
\epsfig{file=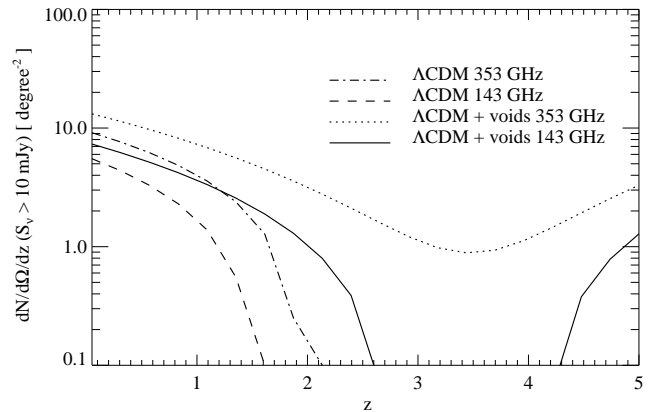,width=9cm}
\caption{Redshift distribution of the sources with flux $S_{143, \:353 \:\tx{GHz}} > 10\; \mJy$ in 
the \lcdm (dashed and dash-dotted lines) and \lcdm + voids (solid and dotted lines) simulations.}
\label{fig:SZDist}
\end{figure}

Fig.~\ref{fig:SZDist} gives the expected redshift distribution of SZ sources with  $S_{\nu}> 10\; \mJy$ 
in the \lcdm + voids and \lcdm models (solid and dotted, dashed 
and dash-dotted lines respectively), for 143 and 353 GHz. The curves have been fitted 
by polynomials in the relevant range. While 
there are virtually no sources with $S_{143,\;353\; \tx{GHz}}> 10\;\mJy$ at $z\ga 1.6$ ($z\ga 2.1$) in $G$, the distribution 
of such objects in the void model is different. At 143 GHz, there are more sources  
in $V$ at low redshift compared to $G$ (a factor 1.3 in excess at $z=0$, reaching 
more than 2 at $z=1$), and the distribution of sources extends to $z=2.6$ on the window shown.  
At $z>4.3$, beyond the ``dip'', the differential number counts of sources in $V$ at 143 GHz again exceed 0.1, 
and keep increasing to $z\sim5$ and earlier. At 353 GHz, SZ sources are intrinsically brighter 
and the differential number counts in $V$ exceed 0.1 all the way to $z\sim5$: 
the number counts decrease from $z=0$ to $z\sim3.4$, then increase to $z=5$ and beyond. 
At $z=0$, there is an excess by a factor 1.6 over the same $V$ counts at 143 GHz, up to $z\sim2$ 
where the ratio increases. At $z=5$ the ratio between the 353 and 143 GHz differential counts in $V$ is 2.5. 
With respect to $G$, the 353 GHz differential number counts in $V$ are higher by 60 percent at $z=0$, 
and by more than an order of magnitude at $z=1.8$. At low redshift $z\sim0$, the excess of sources in $V$ compared to $G$ 
 is accounted for by the excess of clusters of mass $M_{\tx{tot}}\sim10^{14} \hmsun$ seen Fig.~\ref{fig:MF}. 
The slower decrease in the number counts between $z=0$ and $z=3$ in $V$ compared to $G$ is the direct consequence
of the slower late-time evolution of the cluster mass function in $V$ compared to $G$.  
Finally, the increase in the differential number counts at $z\gsim4$ in $V$ is due to the flattening 
and decrease of the angular diameter distance at such redshifts in \lcdmCom together with the presence 
of massive, sufficiently hot haloes at these epochs in the void model. 

In reality, there will be a residual number of sources with $S_{143,\;353\;\tx{GHz}}> 10\; \mJy$ at $z>5$ in the void model, 
as the results shown here provide only a lower limits to the counts, but the difference 
with respect to \lcdm is already a factor of 2 at $z=5$ and SZ surveys may therefore falsify the void model. 
The detection of sufficiently bright high-redshift SZ sources, on the other hand, would be a hint towards non-gaussianity.

\begin{figure}					
\centering
\epsfig{file=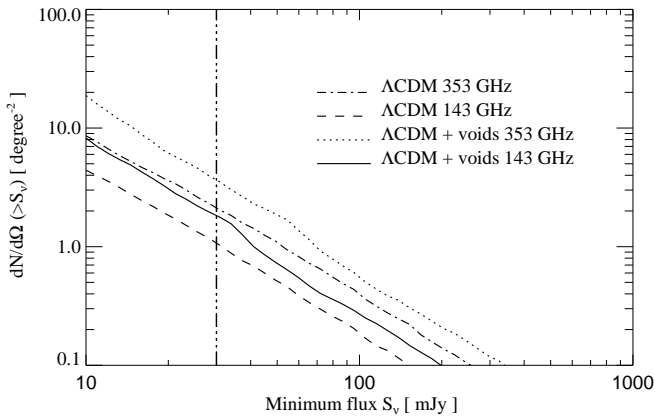,width=9cm}
\caption{Number counts of SZ sources to $z=5$ in the \lcdm (dashed and dash-dotted lines) and 
\lcdm + voids  (solid and dotted lines) models as a function of the minimum flux $S_{\nu,\tx{min}}$ at 143 and 353 GHz. 
Note the  enhancement (factor $\sim2.5$) of the counts above $S_{\nu,\tx{min}}=10\; \mJy$ at 353 GHz in the model with primordial voids. 
The vertical line shows the sensitivity of \emph{Planck}; at this level, number counts differ by a factor $\sim2$.}
\label{fig:SZCountsThresh}
\end{figure}

To see how this conclusion varies with sensitivity, Fig.~\ref{fig:SZCountsThresh} gives 
the number counts of SZ sources expected (up to $z=5$) in the \lcdm (dashed and dash-dotted lines) and 
\lcdm + voids (solid and dotted lines) models as a function of the flux threshold. 
Above a flux limit of 10 mJy, there are  $\sim2$ ($\sim2.5$) times more integrated 
counts at 143 (353) GHz in the \lcdm + voids model compared to \lcdmDot  The sensitivity 
of \emph{Planck} to the SZ effect is of order 30 mJy \citep{Bart01} as shown by 
the vertical dash-triple dotted line, and the corresponding counts are enhanced by a factor 
1.8 (2)  at the two frequencies. For sources brighter than 100 mJy however, the excess 
in the number counts of sources in the primordial void model compared to \lcdm drops 
to a factor 1.5 (1.25) at 143 (353) GHz, a ratio maybe too small to falsify the void model 
against \lcdm with upcoming observations, even if a more detailed computation is needed at this level.  
Among many ground-based examples, the Bonn-Berkeley \emph{APEX - SZ} survey \footnote{\tt http://bolo.berkeley.edu/apexsz/} 
will cover 100 square degrees and is expected to reach a sensitivity of 3 mJy at 150 GHz. This 300-hour experiment will detect $\sim1000$ clusters  
with mass $M_{\tx{tot}}> 2\times10^{14} \hmsun$ up to $z=2$ (assuming a \lcdm model). It will provide strong constraints 
on non-gaussian models like primordial voids. 

\subsection[]{Strong lensing}
\label{sec:Obs:Lens}

The statistics of strong amplification of the images of background sources 
by massive concentrated structures is a powerful probe 
of the cosmological parameters (for example, \citealt{Bart98}).  
Clearly, they will also be directly affected by the primordial 
non-gaussianity considered here.

We assume that only the clusters that we can resolve in the simulations,
($M_{\tx{tot}}\ge M_{\tx{min}}$) produce strong lenses. We model the haloes
with singular isothermal spheres (SIS) \citep{Pea82} and measure their
velocity dispersions directly from the simulations.  \citet{Per02} derive
analytical predictions with NFW profiles for the differential amplification
probability $P(A)$ and compare them to the predictions obtained with SIS
profiles. For sources at $z_{\tx{s}}=4$ and 7, at $A\la 3$ they find their
\lcdm model with NFW profiles to be more efficient than with SIS profiles. At   
$A\ga 7$, SIS profiles yield the higher $P(A)$, but the probabilities for NFW 
and SIS profiles stay within a factor 3 of each other even at larger 
amplifications. We note here that we also neglect the impact of substructure 
and asphericity on the cross section for strong lensing, 
which may be significant but needs better understanding (see, e.g., \citealt{Men02}).

\citet{Bart98} have computed the number of giant arcs expected on the 
whole sky for a series of gaussian CDM models and shown that only open CDM 
models could reproduce the total number of arcs seen in the EMSS sample. In 
particular, their open CDM model produced an order of magnitude more giant 
arcs than their \lcdm model, assuming a redshift $z_{\tx{s}}=1$ for their sources. 
In a recent work, \citet{Wambs03} study the effect of the redshift distribution of sources 
in the predicted number counts of giant arcs in a \lcdm cosmology. 
They show that the lensing optical depth is a very steep function of the source redshift for 
$0.5<z_{\tx{s}}<2.5$ and that it increases further for $z_{\tx{s}}>2.5$.  

In fact, the order of magnitude discrepancy 
between the EMSS counts and the \lcdm prediction is solved 
if one takes a more extended distribution of sources to high redshifts. 
This is more realistic as a large fraction of the giant arcs \citet{Gla03} 
count in the Red-Sequence Cluster Survey (RCS) have high redshifts:  
$1.7< z_{\tx{arc}}< 4.9$.  Putting a third of the sources at $z_{\tx{s}}=1.5$ 
rather than all sources at $z_{\tx{s}}=1$ increases 
the \lcdm predicted counts by a factor 3, and putting 7 percent of the sources 
at $z_{\tx{s}}>3$ increases the counts by a factor of 7, 
bringing \lcdm in agreement with the observations \citep{Wambs03}. 

Because the lensing optical depth is such a steep function of source redshift, 
one expects the large differences between the \lcdm and the primordial voids models 
in the mass function of clusters at $z\sim 1-3$ (probing high source redshifts) 
to significantly affect the number counts of giant arcs, even without classifying 
the counts along arc (source) redshift.  In the following, for simplicity, 
we compute our strong lensing statistics putting all our sources at high redshifts: $z_{\tx{s}}=3$ 
and then at $z_{\tx{s}}=5$. Detailed comparison to observations would require 
a more realistic model for the distribution of source redshifts, but our plots show 
the level of discrepancy expected between \lcdm and \lcdm + voids. 

\begin{figure}					
\centering
\epsfig{file=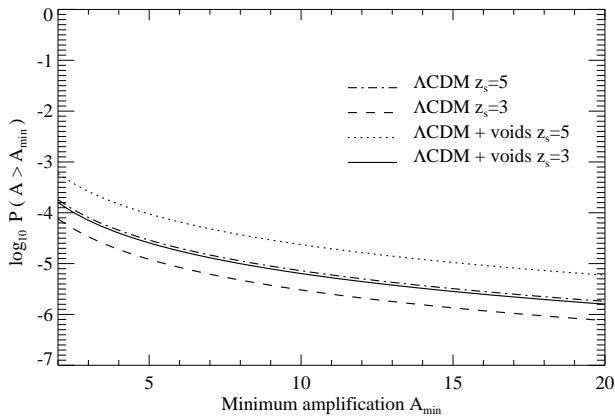,width=9cm}
\caption{Probability $P(A)$ that a line of sight is magnified by $A>A_{\tx{min}}$ for the 
\lcdm (dashed and dash-dotted) and \lcdm + voids model (solid and dotted) for two source redshifts
($z_{\tx{s}}=3$ and 5 respectively). We suppose that the only lenses are the massive DM haloes 
with $M_{\tx{tot}}>M_{\tx{min}}=3.16\times10^{12} \hmsunCom$ modelled as singular isothermal spheres.}
\label{fig:LensProb}
\end{figure}

\begin{figure}					
\centering
\epsfig{file=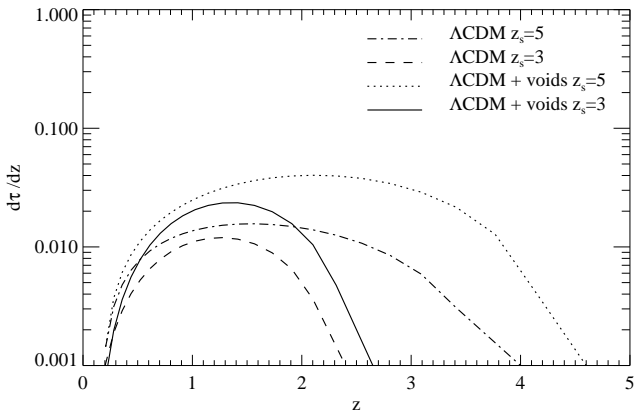,width=9cm}
\caption{Redshift distribution of the optical depth to strong lensing $\tx{d}\tau/\tx{d}z$ contributed by our resolved haloes. 
The legend is the same as for Fig.~\ref{fig:LensProb}. Note the ratio of $\sim2$ to 2.5 between 
the primordial void model and the gaussian \lcdm cosmology.}
\label{fig:TauLens}
\end{figure}

Fig.~\ref{fig:LensProb} shows the cumulative probability $P(A>A_{\tx{min}})$
that a line of sight has an amplification larger than $A_{\tx{min}}$ in the
\lcdm and \lcdm + voids cosmologies and for the two values of $z_{\tx{s}}$.  
We take the strong lensing regime to be $A\ga2$.  At $A_{\tx{min}}\ga 10$, the cumulative
probability for strong lensing is a factor $\sim 2$ ($\sim 4$) higher in
the void model than in the \lcdm case, for $z_{\tx{s}}=3$ (5). For sources at $z_{\tx{s}}=5$, 
we note that the enhancement is also larger than the possible bias due to our 
choice of an SIS rather than NFW halo profile. We conclude that the optical depth to strong lensing   
 is increased in the \lcdm + voids model compared to \lcdmCom and that the number 
counts of giant arcs may differ by up to a factor of 4, hence overpredicting the observations compared to \lcdmDot 
The optical depth to high-redshift galaxies will be increased 
by a factor of 4 in the void model, and the 
the number counts of high-redshift giant arcs (with $z_{\tx{arc}}\sim5$) 
will be increased by \emph{more} than a factor of 4 in the \lcdm + voids model compared to the \lcdm model, 
because the comoving density of bright ($M_{\tx{*}}> 10^{10}$ \hmsunKet 
galaxies is also higher at $z\sim5$ in the  \lcdm + voids model than in the \lcdm model. 
(We verified this last point using a semi-analytical model for galaxy formation 
similar to that described in \citealt{Ma01}). If the density of the lensed population evolves similarly 
in the \lcdm and the \lcdm + voids models, then the mean redshift of the lenses will be shifted 
to higher values in $V$ compared to $G$: Fig.~\ref{fig:TauLens} shows the redshift distribution of the 
total optical depth $\tx{d}\tau/\tx{d}z$ to strong lensing as defined in section 4.2 of \citet{Pea99Book}, 
for $z_{\tx{s}}=3$ and 5. The total optical depth $\tau$ is larger in the $V$ than in the $G$ simulation 
(note the factors $\sim2$ and $\sim2.5$  decrease between the $V$ and $G$ curves at $z_{\tx{s}}=3$ and 5 in Fig.~\ref{fig:TauLens}). 


Finally, we have assumed initially compensated voids surrounded by a thin shell growing as the 
underdensity expands in comoving coordinates.  
This large-scale configuration could  \emph{per se} constitute an efficient lens. 
However, \citet{Am99} show that only voids with radius larger than $\sim100$ \hMpc today 
induce weak gravitational lensing with a signal to noise ratio 
greater than unity in observables like color-dependent density 
magnification or aperture densitometry. 
Even for strong underdensities like $\delta_{\tx{void}}\sim-1$, none of these weak 
lensing techniques will be able to find a significant signature of the voids considered here.


\section[]{Conclusions}
\label{sec:CCL}

We have simulated cluster formation in a physically plausible non-gaussian primordial void model where empty and fully compensated bubbles 
surviving from inflation  together with the gaussian adiabatic  CDM-type perturbations 
provide the seeds for the development of structure. This model shares the cosmological parameters of \lcdm and 
possesses gaussian statistics on large scales while non-gaussianity only affects the one-point distribution on cluster scales.   
It is an attractive alternative to \lcdm as it may explain the excess of CMB temperature anisotropy 
power at $l\sim2500$ recently observed by CBI and as it could account for the large voids seen in the nearby galaxy surveys \citep{Gri03}. 
While analysis of high-resolution CMB maps using higher-order moments will provide further constraints 
on primordial non-gaussianity, we have shown the evolution of the cluster mass function to be a strong constraint at low and intermediate redshifts.

At startup, the total energy of the network of voids and shells in our simulation   
is comparable to the potential energy of the gaussian fluctuations.  
Even if strong underdensities are present on Mpc scales very early on compared to \lcdmCom 
 we have found that our $200\; \hMpc$-side, 
$128^3$-particle simulations which start shortly after recombination can provide a reliable  
estimate for the cluster mass function of the non-gaussian model. 

The power spectrum of the void model measured after setting the initial conditions shows a strong feature 
characteristic of the voids at $k\gsim 0.1$ \hMpcInvCom which is then erased during 
the evolution by the non-gaussian initial conditions. The $z=0$ matter power spectra of the \lcdm + void model 
is close to that of the gaussian model, with little additional small-scale power at $k\gsim0.5\, \hMpcInvDot$  
The $z=0$ mass function  gives a similar number of massive clusters ($M_{\tx{tot}}>4\times10^{14}\hmsunKet$ 
in the \lcdm + voids model and in the concordance \lcdm scenario. Our high cluster 
abundance compared to observations is a consequence of the \emph{WMAP}+SDSS normalisation.  
The evolution of the cluster mass function up to $z\sim~3$ in the void scenario differs strongly  
 from that of the \lcdm model and is much more ``efficient'' than the sole $z=0$ 
mass function in distinguishing between the two. The substantial evolution in the cluster baryon mass function seen by  
\citet{Vik03} between $z=0$ and $z\sim0.5$ needs confirmation before we can falsify the non-gaussian 
model which has little late-time evolution. 
 
In fact, better constraints on the void model are obtained from two other statistics at $z\gsim1.5$. 
We have first shown that the integrated number counts of SZ sources is higher 
by a factor 2.5 (resp. 2) in the \lcdm + voids model compared to the \lcdm model, 
for 353 GHz flux greater than 10 mJy (resp. 30 mJy, the resolution of Planck). 
We have then used the optical depth to strong gravitational lensing as another   
possible discriminant between the primordial void model and \lcdmDot 
We have shown that the number counts of high-redshift, $z\gsim3$ (resp. 5) 
arcs is expected to be more abundant by a factor of 2 (resp. 4) in the non-gaussian scenario, 
for the same underlying lensed population. Because 
the optical depth to strong lensing is a steep function of redshift, 
we expect the total number of giant arcs observed to 
be increased by a factor $\gsim2$ in the \lcdm + voids model compared to \lcdmDot  
As a result, the \lcdm + void scenario overpredicts the number counts of giant arcs 
seen in the EMSS compared to the concordance \lcdmDot  

An additional contribution to the CMB power spectrum at high $\ell$ will
also be generated by the thermal SZ contribution from unresolved clusters.
The joint limits on a non-gaussian contribution from clusters to the power
 spectrum, including 
the ROSAT/WMAP cross-correlation constraints 
\citep{Die03b} and  the CBI and ACBAR 
experiments, will be presented elsewhere. 
 
Not only do such simple tests give the opportunity to rule out a particular  
set of models, they also more generally bring out simple constraints that have  
to be satisfied (e.g. with numerical simulations) as one proposes non-gaussian alternatives to the current paradigm. 



\section*{Acknowledgements}
\label{sec:Ack}

We thank the referee for a number of remarks 
which significantly improved the manuscript. HM is supported by PPARC. 

\bsp
 
\label{lastpage}

\bibliographystyle{mnras}



\end{document}